\documentclass[preprint]{emulateapj} 

\usepackage{natbib}
\usepackage{hyperref}
\hypersetup{colorlinks,citecolor=Blue,linkcolor=Red,urlcolor=Blue}
\usepackage{amsmath}
\usepackage{empheq}
\usepackage[usenames,dvipsnames]{color}
\allowdisplaybreaks 

\newcommand{\icarus}{Icarus}

\begin{document}
 
\title{Evidence for a Distant Giant Planet in the Solar System}  
\author{Konstantin Batygin \& Michael E. Brown} 

\affil{Division of Geological and Planetary Sciences, California Institute of Technology, Pasadena, CA 91125} 
\email{kbatygin@gps.caltech.edu}
 
\newcommand{\Ham}{\mathcal{H}}
\newcommand{\G}{\mathcal{G}}
\newcommand{\appropto}{\mathrel{\vcenter{\offinterlineskip\halign{\hfil$##$\cr\propto\cr\noalign{\kern2pt}\sim\cr\noalign{\kern-2pt}}}}}

\begin{abstract} 

Recent analyses have shown that distant orbits within the scattered disk population of the Kuiper belt exhibit an unexpected clustering in their respective arguments of perihelion. While several hypotheses have been put forward to explain this alignment, to date, a theoretical model that can successfully account for the observations remains elusive. In this work we show that the orbits of distant Kuiper belt objects cluster not only in argument of perihelion, but also in physical space. We demonstrate that the perihelion positions and orbital planes of the objects are tightly confined and that such a clustering has only a probability of 0.007\% to be due to chance, thus requiring a dynamical origin. We find that the observed orbital alignment can be maintained by a distant eccentric planet with mass $\gtrsim10\,m_{\oplus}$ whose orbit lies in approximately the same plane as those of the distant Kuiper belt objects, but whose perihelion is 180 degrees away from the perihelia of the minor bodies. In addition to accounting for the observed orbital alignment, the existence of such a planet naturally explains the presence of high perihelion Sedna-like objects, as well as the known collection of high semimajor axis objects with inclinations between 60 and 150 degrees whose origin was previously unclear. Continued analysis of both distant and highly inclined outer solar system objects provides the opportunity for testing our hypothesis as well as further constraining the orbital elements and mass of the distant planet.

\end{abstract} 

\maketitle

\section{Introduction} \label{sect1}

The recent discovery of 2012$\,$VP113, a Sedna-like body and a potential additional member of the inner Oort cloud, prompted \citet{TrujilloSheppard} to note that a set of Kuiper belt objects (KBOs) in the distant solar system exhibits unexplained clustering in orbital elements. Specifically, objects with perihelion distance larger than the orbit of Neptune and semi-major axis greater than $150\,$AU  -- including 2012$\,$VP113 and Sedna --  have arguments of perihelia, $\omega$, clustered approximately around zero. A value of $\omega = 0$ requires that the object's perihelion lies precisely at the ecliptic, and during ecliptic-crossing the object moves from south to north (i.e. intersects the ascending node). While observational bias does preferentially select for objects with perihelia (where they are closest and brightest) at the heavily observed ecliptic, no possible bias could select only for objects moving from south to north. Recent simulations \citep{delafuetes_forever} confirmed this lack of bias in the observational data. The clustering in $\omega$ therefore appears to be real.

Orbital grouping in $\omega$ is surprising because gravitational torques exerted by the giant planets are expected to randomize this parameter over the multi-Gyr age of the solar system. In other words, the values of $\omega$ will not stay clustered unless some dynamical mechanism is currently forcing the alignment. To date, two explanations have been proposed to explain the data.

\citet{TrujilloSheppard} suggest that an external perturbing body could allow $\omega$ to librate about zero via the \textit{Kozai mechanism}\footnote{Note that the invoked variant of the Kozai mechanism has a different phase-space structure from the Kozai mechanism typically discussed within the context of the asteroid belt (e.g. \citealt{ThomasMorbidelli1996}).}. As an example, they demonstrate that a 5 Earth mass body on a circular orbit at $210\,$AU can drive such libration in the orbit of 2012$\,$VP113. However, \citet{delafuetes_forever} note that the existence of librating trajectories around $\omega = 0$ requires the ratio of the object to perturber semi-major axis to be nearly unity. This means that that trapping all of the distant objects within the known range of semi-major axes into Kozai resonances likely requires multiple planets, finely tuned to explain the particular data set. 

\begin{figure*}[t]
\includegraphics[width=\textwidth]{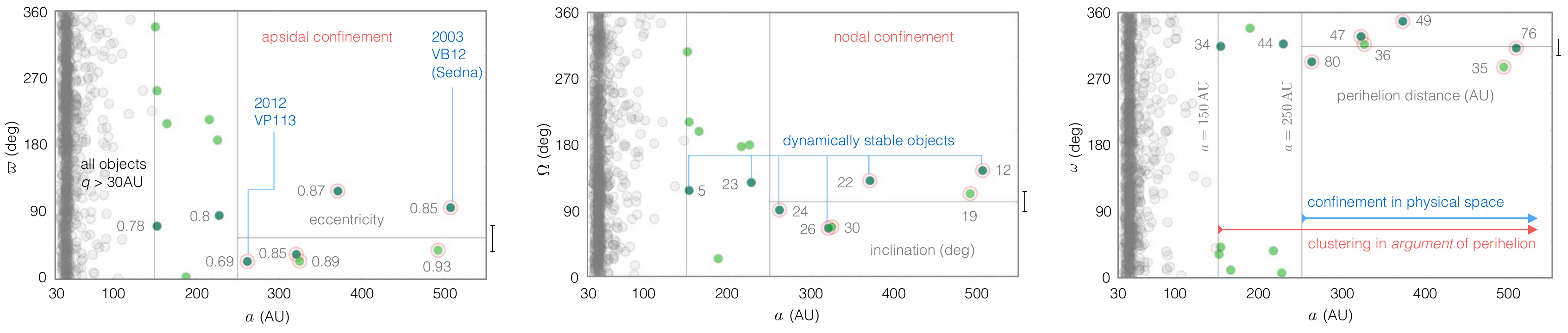}
\caption{Orbits of well-characterized Kuiper-belt objects with perihelion distances in excess of $q>30\,$AU. The left, middle, and right panels depict the longitude of perihelion, $\varpi$, longitude of ascending node, $\Omega$, and argument of perihelion $\omega$ as functions of semi-major axes. The orbits of objects with $a<150\,$AU are randomly oriented and are shown as gray points. The argument of perihelion displays clustering beyond $a>150\,$AU, while the longitudes of perihelion and ascending node exhibit confinement beyond $a>250\,$AU. Within the $a>150\,$AU subset of objects, dynamically stable bodies are shown with blue-green points, whereas their unstable counterparts are shown as green dots. By and large, the stable objects are clustered in a manner that is consistent with the $a>250\,$AU group of bodies. The eccentricities, inclinations and perihelion distances of the stable objects are additionally labeled. The horizontal lines beyond $a>250\,$AU depict the mean values of the angles and the vertical error bars depict the associated standard deviations of the mean.}
\label{Fig1}
\end{figure*}

Further problems may potentially arise with the Kozai hypothesis. \citet{TrujilloSheppard} point out that the Kozai mechanism allows libration about both $\omega = 0$ as well as $\omega = 180$, and the lack of $\omega \sim 180$ objects suggests that some additional process originally caused the objects to obtain $\omega \sim 0$. To this end, they invoke a strong stellar encounter to generate the desired configuration. Recent work \citep{Jilkova2015} shows how such an encounter could in principle lead to initial conditions that would be compatible with this narrative. Perhaps a greater difficulty lies in that the dynamical effects of such a massive perturber might have already been visible in the inner solar system. \citet{Iorio2014} has analyzed the effects of a distant perturber on the precession of the apsidal lines of the inner planets and suggests that, particularly for low inclination perturbers, objects more massive than the Earth with $a \sim 200-300\,$AU are ruled out from the data (see also \citealt{Iorio2012}). 

As an alternative explanation, \citet{MadiganMcCourt2015} have proposed that the observed properties of the distant Kuiper belt can be attributed to a so-called \textit{inclination instability}. Within the framework of this model, an initially axisymmetric disk of eccentric planetesimals is reconfigured into a cone-shaped structure, such that the orbits share an approximately common value of $\omega$ and become uniformly distributed in the longitude of ascending node, $\Omega$. While an intriguing proposition, additional calculations are required to assess how such a self-gravitational instability may proceed when the (orbit-averaged) quadrupolar potential of the giant planets, as well as the effects of scattering are factored into the simulations. Additionally, in order to operate on the appropriate timescale, the inclination instability requires $1-10$ Earth masses of material to exist between $\sim100\,$AU and $\sim 10,000\,$AU \citep{MadiganMcCourt2015}.

Such an estimate is at odds with the negligibly small mass of the present Sedna population \citep{Schwamb2010}. To this end, it is worth noting that although the \textit{primordial} planetesimal disk of the solar system likely comprised tens of Earth masses \citep{Tsiganis2005,Levison2008,Levison2011,Batygin2011}, the vast majority of this material was ejected from the system by close encounters with the giant planets, during, and immediately following the transient dynamical instability that shaped the Kuiper belt in the first place. The characteristic timescale for depletion of the primordial disk is likely to be short compared with the timescale for the onset of the inclination instability \citep{Nesvorny2015}, calling into question whether the inclination instability could have actually proceeded in the outer solar system.

In light of the above discussion, here we reanalyze the clustering of the distant objects and propose a different perturbation mechanism, stemming form a single, long-period object. Remarkably, our envisioned scenario brings to light a series of potential explanations for other, seemingly unrelated dynamical features of the Kuiper belt, and presents a direct avenue for falsification of our hypothesis. The paper is organized as follows. In section \ref{sect2}, we reexamine the observational data and identify the relevant trends in the orbital elements. In section \ref{sect3}, we motivate the existence of a distant, eccentric perturber using secular perturbation theory. Subsequently, we engage in numerical exploration in section \ref{sect4}. In section \ref{sect5}, we perform a series of simulations that generate synthetic scattered disks. We summarize and discuss the implications of our results in sections \ref{sect6} and \ref{sect7} respectively.

\begin{figure*}[t]
\includegraphics[width=\textwidth]{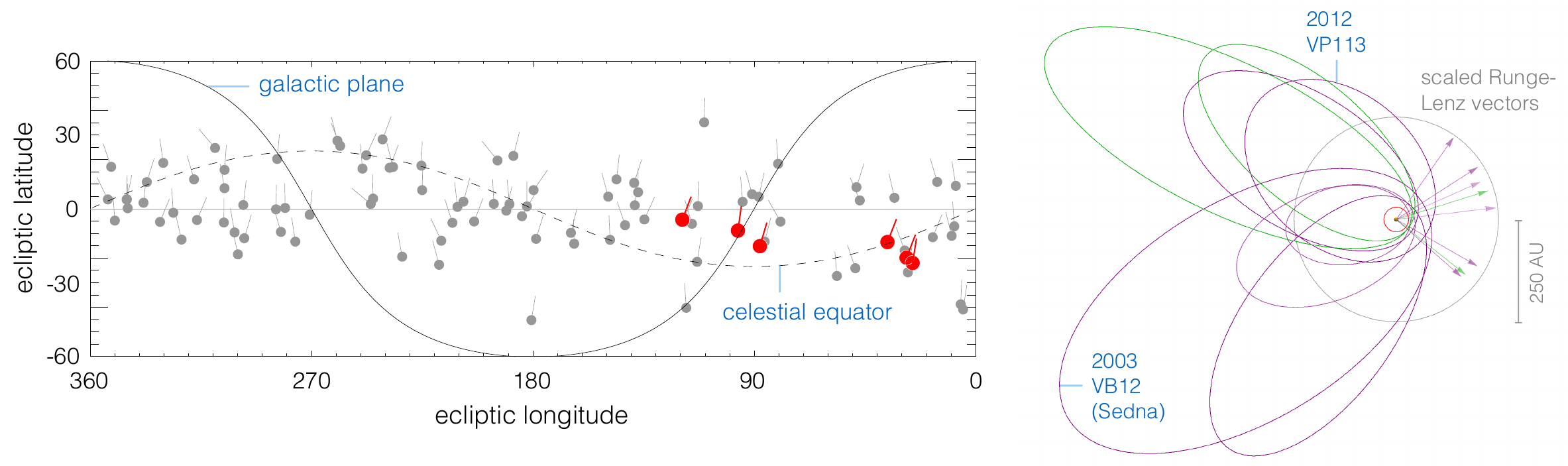}
\caption{Orbital clustering in physical space. The right panels depicts the side and top views of the Keplerian trajectories of all bodies with $a>250\,$AU as well as dynamically stable objects with $a>150\,$AU. The adopted color scheme is identical to that employed in Figure (\ref{Fig1}), and the two thin purple orbits correspond to stable bodies within the $150<a<250\,$AU range. For each object, the directions of the angular momentum and Runge-Lenz (eccentricity) vectors are additionally shown. The left panel shows the location of perihelia of the minor bodies with $q>30\,$AU and $a>50\,$AU on the celestial sphere as points, along with the projection of their orbit poles with adjacent lines. The orbits with $a>250\,$AU are emphasized in red. The physical confinement of the orbits is clearly evident in both panels.}
\label{Fig2}
\end{figure*}

\section{Orbital element analysis} \label{sect2}
In their original analysis, \citet{TrujilloSheppard} examined $\omega$ as a function of semi-major axis for all objects with perihelion, $q$, larger than Neptune's orbital distance (Figure \ref{Fig1}). They find that all objects with $q>30\,$AU and $a>150\,$AU are clustered around $\omega \sim 0$. Excluding objects with $q$ inside of Neptune's orbit is sensible, since an object which crosses Neptune's orbit will be influenced by recurrent close encounters. However, many objects with $q>30\,$AU can also be destabilized as a consequence of Neptune's overlapped outer mean-motion resonances (e.g. \citealt{MorbyBook}), and a search for orbits that are not contaminated by strong interactions with Neptune should preferably exclude these objects as well.

In order to identify which of the $q>30 $AU and $a>150\,$AU KBOs are strongly influenced by Neptune, we numerically evolved 6 clones of each member of the clustered population for 4 Gyr. If more than a single clone in the calcuations exhibited large-scale semi-major axis variation, we deemed such an objects dynamically unstable\footnote{In practice, large-scale orbital changes almost always result in ejection.}. Indeed, many of the considered KBOs (generally those with $30<q<36\,$AU) experience strong encounters with Neptune, leaving only 6 of the 13 bodies largely unaffected by the presence of Neptune. The stable objects are shown as dark blue-green dots in Figure (\ref{Fig1}) while those residing on unstable orbits are depicted as green points. 

Interestingly, the stable objects cluster not around $\omega = 0$ but rather around $\omega = 318 \pm 8\, \deg$, grossly inconsistent with the value predicted from by the Kozai mechanism. Even more interestingly, a corresponding analysis of longitude of ascending node, as a function of the semi-major axis reveals a similarly strong clustering of these angles about $\Omega = 113 \pm 13\, \deg$ (Figure \ref{Fig1}). Analogously, we note that \textit{longitude} of perihelion\footnote{Unlike the argument of perihelion, $\omega$, which is measured from the ascending node of the orbit, the longitude of perihelion, $\varpi$, is an angle that is defined in the inertial frame.}, $\varpi = \omega + \Omega$, is grouped around $\varpi = 71 \pm 16\,\deg$. Essentially the same statistics emerge even if long-term stability is disregarded but the semi-major axis cut is drawn at $a=250\,$AU. The clustering of both $\varpi$ and of $\Omega$ suggests that not only do the distant KBOs cross the ecliptic at a similar phase of their elliptical trajectories, \textit{the orbits are physically aligned}. This alignment is evident in the right panel of Figure (\ref{Fig2}), which shows a polar view of the Keplerian trajectories in inertial space. 

To gauge the significance of the physical alignment, it is easier to examine the orbits in inertial space rather than orbital element space. To do so, we calculate the location of the point of perihelion for each of the objects and project these locations into ecliptic coordinates\footnote{The vector joining the Sun and the point of perihelion, with a magnitude $e$ is formally called the Runge-Lenz vector.}. In addition, we calculate the pole orientation of each orbit and project it onto the plane of the sky at the perihelion position. The left panel of Figure (\ref{Fig2}) shows the projected perihelion locations and pole positions of all known outer solar system objects with $q>30\,$AU and $a>50\,$AU. The 6 objects with $a>250\,$AU, highlighted in red, all come to perihelion below the ecliptic and at longitudes between $20\,\deg$ and $130\deg$. 

Discovery of KBOs is strongly biased by observational selection effects which are poorly calibrated for the complete heterogeneous Kuiper belt catalog. A clustering in perihelion position on the sky could be caused, for example, by preferential observations in one particular location. The distribution of perihelion positions across the sky for all objects with $q>30$ and $a>50\,$AU appears biased toward the equator and relatively uniform in longitude. No obvious bias appears to cause the observed clustering. In addition, each of our 6 clustered objects was discovered in a separate survey with, presumably, uncorrelated biases. 

We estimate the statistical significance of the observed clustering by assuming that the detection biases for our clustered objects are similar to the detection biases for the collection of all objects with $q>30\,$AU and $a>50\,$AU. We then randomly select 6 objects from the sample 100,000 times and calculate the root-mean-square (RMS) of the angular distance between the perihelion position of each object and the average perihelion position of the selected bodies. Orbits as tightly clustered in perihelion position as the 6 observed KBOs occur only 0.7\% of the time.  Moreover, the objects with clustered perihelia also exhibit clustering in orbital pole position, as can be seen by the nearly identical direction of their projected pole orientations. We similarly calculated the RMS spread of the polar angles, and find that a cluster as tight as that observed in the data occurs only 1\% of the time. The two measurements are statistically uncorrelated, and we can safely multiply the probabilities together to find that the joint probability of observing both the clustering in perihelion position and in pole orientation simultaneously is only 0.007\%. Even with only 6 objects currently in the group, the significance level is about 3.8$\,\sigma$. It is extremely unlikely that the objects are so tightly confined due purely to chance.

Much like confinement in $\omega$, orbital alignment in physical space is difficult to explain because of differential precession. In contrast to clustering in $\omega$, however, orbital confinement in physical space cannot be maintained by either the Kozai effect or the inclination instability. This physical alignment requires a new explanation. 

\section{Analytical Theory} \label{sect3}

Generally speaking, coherent dynamical structures in particle disks can either be sustained by self-gravity \citep{Tremaine1998,Touma2009} or by gravitational shepherding facilitated by an extrinsic perturber \citep{GoldreichTremaine1982,Chiang2009}. As already argued above, the current mass of the Kuiper belt is likely insufficient for self-gravity to play an appreciable role in its dynamical evolution. This leaves the latter option as the more feasible alternative. Consequently, here we hypothesize that the observed structure of the Kuiper belt is maintained by a gravitationally bound perturber in the solar system.

To motivate the plausibility of an unseen body as a means of explaining the data, consider the following analytic calculation. In accord with the selection procedure outlined in the preceding section, envisage a test particle that resides on an orbit whose perihelion lies well outside Neptune's orbit, such that close encounters between the bodies do not occur. Additionally, assume that the test particle's orbital period is not commensurate (in any meaningful low-order sense - e.g. \citealt{NesvornyRoig2001}) with the Keplerian motion of the giant planets. 

\begin{figure*}[t]
\includegraphics[width=\textwidth]{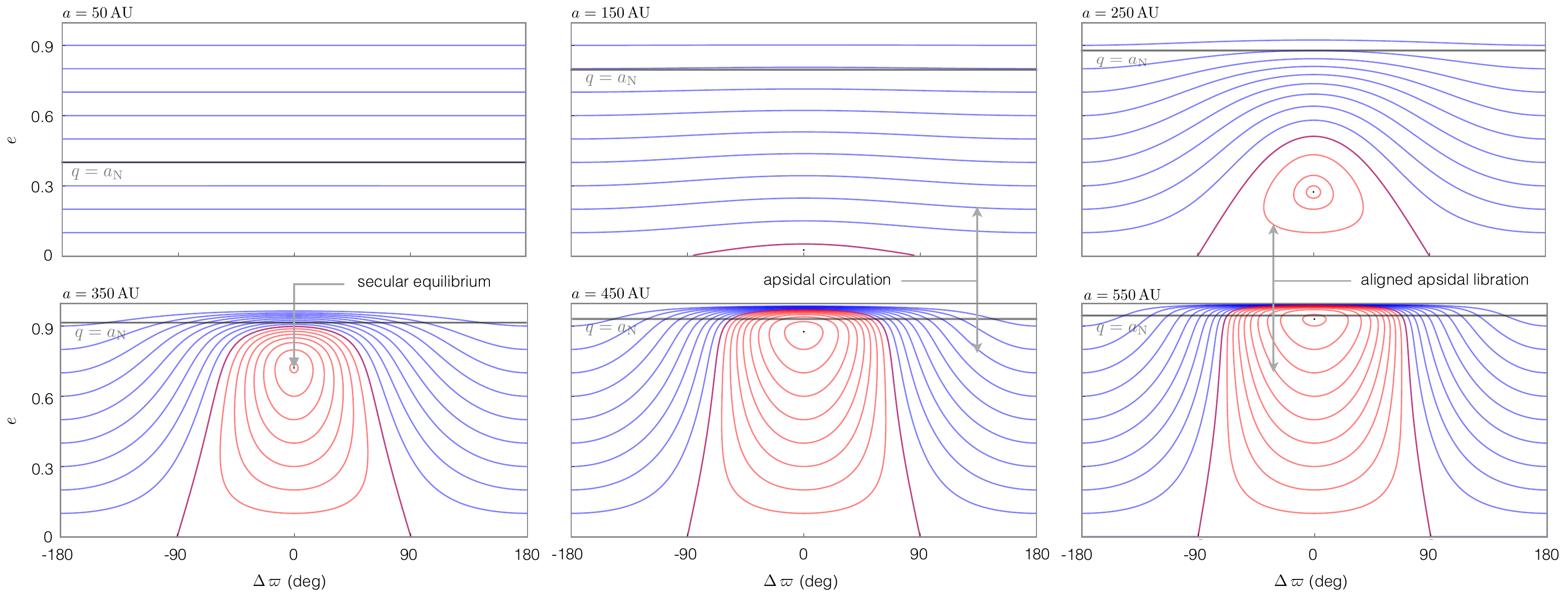}
\caption{Phase-space portraits (projected into orbital element space) corresponding to the autonomous Hamiltonian (\ref{H3}). Note that unlike Figure (\ref{Fig1}), here the longitude of perihelion (plotted along the x-axis) is measured with respect to the apsidal line of the perturber's orbit. Red curves represent orbits that exhibit apsidal libration, whereas blue curves denote apsidal circulation. On each panel, the eccentricity corresponding to a Neptune-hugging orbit is emphasized with a gray line. The unseen body is assumed to have a mass of $m'=10m_{\oplus}$, and reside on a $a'=700\,$AU, $e'=0.6$ orbit.}
\label{Fig3}
\end{figure*}

The long-term dynamical behavior of such an object can be described within the framework of secular perturbation theory \citep{Kaula1964}. Employing Gauss's averaging method (see Ch. 7 of \citealt{MD99,Touma2009}), we can replace the orbits of the giant planets with massive wires and consider long-term evolution of the test particle under the associated torques. To quadrupole order\footnote{The octupolar correction to equation (\ref{H1}) is proportional to the minuscule eccentricities of the giant planets, and can safely be neglected.} in planet-particle semi-major axis ratio, the Hamiltonian that governs the planar dynamics of the test particle is:
\begin{align}
\label{H1}
\Ham = -\frac{1}{4}\frac{\mathcal{G} M}{a} \left(1-e^2 \right)^{-3/2} \sum_{i=1}^{4} \frac{m_i a_i^2}{M a^2}.
\end{align}
In the above expression, $\G$ is the gravitational constant, $M$ is the mass of the Sun, $m_i$ and $a_i$ are the masses and semi-major axes of the giant planets, while $a$ and $e$ are the test particle's semi-major axis and eccentricity.

Equation (\ref{H1}) is independent of the orbital angles, and thus implies (by application of Hamilton's equations) apsidal precession at constant eccentricity with the period\footnote{Accounting for finite inclination of the orbit enhances the precession period by a factor of $1/\cos^2(i) \simeq 1+i^2 + ...$ .}:
\begin{align}
\label{Precession}
\frac{\mathcal{P}_{\omega}}{\mathcal{P}} = \frac{4}{3} \left(1-e^2 \right)^2\sum_{i=1}^{4} \frac{M a^2}{m_i a_i^2},
\end{align}
where $\mathcal{P}$ is the orbital period. As already mentioned above, in absence of additional effects, the observed alignment of the perihelia could not persist indefinitely, owing to differential apsidal precession. As a result, additional perturbations (i.e. harmonic terms in the Hamiltonian) are required to explain the data. 

Consider the possibility that such perturbations are secular in nature, and stem from a planet that resides on a planar, exterior orbit. Retaining terms up to octupole order in the disturbing potential, the Hamiltonian takes the form \citep{Mardling2013}:
\begin{align}
\label{H2}
\Ham &= -\frac{1}{4}\frac{\mathcal{G} M}{a} \left( 1-e^2 \right)^{-3/2} \sum_{i=1}^{4} \frac{m_i a_i^2}{M a^2} \nonumber \\
&- \frac{\mathcal{G}\, m'}{a'} \bigg[ \frac{1}{4} \left( \frac{a}{a'} \right)^2 \frac{1+3\,e^2/2}{\big(1-(e')^2\big)^{3/2}} \nonumber \\
&- \frac{15}{16} \left( \frac{a}{a'} \right)^3 e \, e' \frac{1+3\,e^2/4}{\big(1-(e')^2\big)^{5/2}} \cos(\varpi' - \varpi) \bigg],
\end{align}
where primed quantities refer to the distant perturber (note that for planar orbits, longitude and argument of perihelion are equivalent). Importantly, the strength of the harmonic term in equation (\ref{H2}) increases monotonically with $e'$. This implies that in order for the perturbations to be consequential, the companion orbit must be appreciably eccentric.

Assuming that the timescale associated with secular coupling of the giant planets is short compared with the characteristic timescale for angular momentum exchange with the distant perturber (that is, the interactions are \textit{adiabatic} - see e.g. \citealt{Neishtadt1984,BeckerBatygin2013}), we may hold all planetary eccentricities constant and envision the apse of the perturber's orbit to advance linearly in time: $\varpi' = \nu\, t$, where the rate $\nu$ is obtained from equation (\ref{Precession}).

Transferring to a frame co-precessing with the apsidal line of the perturbing object through a canonical change of variables arising from a type-2 generating function of the form $\mathcal{F}_2 = \Phi (\nu \, t - \varpi)$, we obtain an autonomous Hamiltonian \citep{Goldstein}:
\begin{align}
\label{H3}
\Ham &= -\frac{1}{4}\frac{\mathcal{G} M}{a} \left( 1-e^2 \right)^{-3/2} \sum_{i=1}^{4} \frac{m_i a_i^2}{M a^2} \nonumber \\
&+\nu\, \sqrt{\mathcal{G} \, M \, a} \left( 1-  \sqrt{1-e^2} \right) \nonumber \\
&- \frac{\mathcal{G}\, m'}{a'} \bigg[ \frac{1}{4} \left( \frac{a}{a'} \right)^2 \frac{1+3\,e^2/2}{\big(1-(e')^2\big)^{3/2}} \nonumber \\
&- \frac{15}{16} \left( \frac{a}{a'} \right)^3 e \, e' \frac{1+3\,e^2/4}{\big(1-(e')^2\big)^{5/2}} \cos(\Delta\varpi) \bigg],
\end{align}
where $\Phi=\sqrt{\G M a} \left(1 - \sqrt{1-e^2} \right)$ is the action conjugate to the angle $\Delta\varpi = \varpi'-\varpi$. Given the integrable nature of $\Ham$, we may inspect its contours as a way to quantify the orbit-averaged dynamical behavior of the test-particle. 

Figure (\ref{Fig3}) shows a series of phase-space portraits\footnote{Strictly speaking, Figure (\ref{Fig3}) depicts a projection of the phase-space portraits in orbital element space, which is not canonical. However, for the purposes of this work, we shall loosely refer to these plots as phase-space portraits, since their information content is identical.} of the Hamiltonian (\ref{H3}) for various test particle semi-major axes and perturber parameters of $m'=10\,m_{\oplus}$, $a'=700\,$AU and $e'=0.6$. Upon examination, an important feature emerges within the context of our simple model. For test particle semi-major axes exceeding $a\gtrsim200\,$AU, phase-space flow characterized by libration of $\Delta \varpi$ (shown as red curves) materializes at high eccentricities. 

\begin{figure*}[t]
\includegraphics[width=\textwidth]{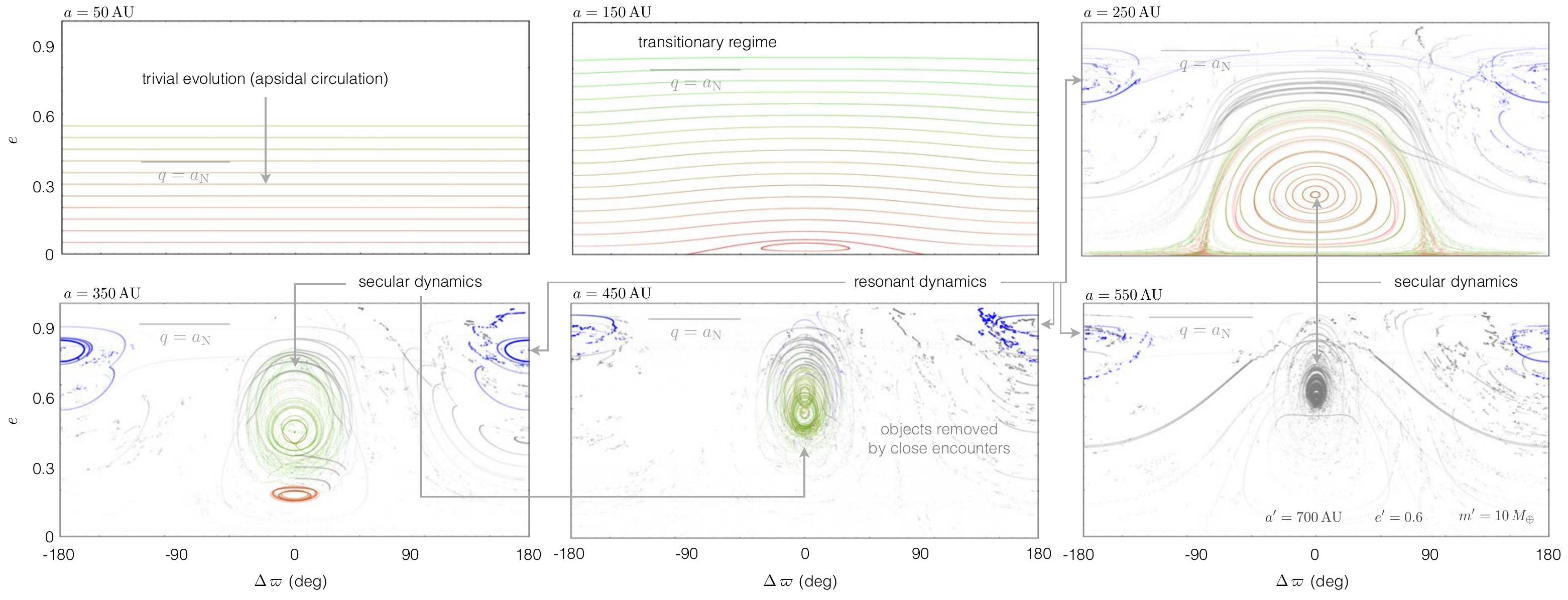}
\caption{Projection of dynamical evolution computed within the framework of N-body simulations into $e-\Delta\varpi$ space. Orbits whose secular evolution facilitates libration of the longitude of perihelion (akin to that depicted with red curves shown in Figure \ref{Fig3}) are shown as orange lines, where the shade is used as a proxy for starting eccentricity (evident in panels corresponding to $a=50,150\,$AU). Long-term unstable orbits are plotted with gray points. Dynamically long-lived trajectories characterized by apsidal anti-alignment are shown with blue points. As discussed in text, these anti-aligned configurations likely derive their dynamical structure and stability from high-order mean-motion commensurabilities associated with the Keplerian motion of the distant planet.}
\label{Fig4}
\end{figure*}

This is consequential for two (inter-related) reasons. First, the existence of a libration island\footnote{Note that Figure (\ref{Fig3}) does not depict any homoclinic curves, so libration of $\Delta \varpi$ is strictly speaking \textit{not} a secular resonance \citep{HenrardLamaitre1983}.} demonstrates that an eccentric perturber can modify the orbital evolution of test particles in such a way as to dynamically maintain apsidal alignment. Second, associated with the libration of $\Delta \varpi$ is modulation of orbital eccentricity. This means that a particle initially on a Neptune-hugging orbit can become detached by a secular increase in perihelion distance. 

We note that the transition from trivial apsidal circulation to a picture where librating trajectories occupy a substantial fraction of the parameter space inherent to Hamiltonian (\ref{H3}) depends sensitively on the employed parameters. In particular, the phase-space portraits exhibit the most dramatic dependence on $a'$, since the harmonic term in equation (\ref{H3}) arises at a higher order in the expansion of the disturbing potential than the term responsible for coupling with the giant planets. Meanwhile, the sensitivity to $e'$ is somewhat diminished, as it dominantly regulates the value of $e$ that corresponds to the elliptic equilibrium points shown in Figure (\ref{Fig3}). On the other hand, in a regime where the last two terms in equation (\ref{H3}) dominate (i.e. portraits corresponding to $a\gtrsim350\,$AU and companions with $a'\sim 500-1000\,$AU, $e'\gtrsim 0.6$ and $m'\gtrsim$ a few Earth masses), $m'$ only acts to determine the timescale on which secular evolution proceeds. Here, the choice of parameters has been made such that the resulting phase-space contours match the observed behavior, on a qualitative level.

Cumulatively, the presented results offer credence to the hypothesis that the observed structure of the distant Kuiper belt can be explained by invoking perturbations from an unseen planetary mass companion to the solar system. Simultaneously, the suggestive nature of the results should be met with a healthy dose of skepticism, given the numerous assumptions made in the construction of our simple analytical model. In particular, we note that a substantial fraction of the dynamical flow outlined in phase-space portraits (\ref{Fig3}) characterizes test-particle orbits that intersect that of the perturber (or Neptune), violating a fundamental assumption of the employed secular theory. 

Moreover, even for orbits that do not cross, it is not obvious that the perturbation parameter $(a/a')$ is ubiquitously small enough to warrant the truncation of the expansion at the utilized order. Finally, Hamiltonian (\ref{H3}) does not account for possibly relevant resonant (and/or short-periodic) interactions with the perturber. Accordingly, the obtained results beg to be re-evaluated within the framework of a more comprehensive model. 

\section{Numerical Exploration} \label{sect4}

In an effort to alleviate some of the limitations inherent to the calculation performed above, let us abandon secular theory altogether and employ direct N-body simulations\footnote{For the entirety of the N-body simulation suite, we utilized the \texttt{mercury6} gravitational dynamics software package \citep{Chambers}. The hybrid symplectic-Bulisch-Stoer algorithm was employed throughout, and the timestep was set to a twentieth of the shortest dynamical timescale (e.g. orbital period of Jupiter).}. For a more illuminating comparison among analytic and numeric models, it is sensible to introduce complications sequentially. In particular, within our first set of numerical simulations, we accounted for the interactions between the test particle and the distant companion self-consistently, while treating the gravitational potential of the giant planets in an orbit-averaged manner. 

Practically, this was accomplished by considering a central object (the Sun) to have a physical radius equal to that of Uranus's semi-major axis ($\mathcal{R} = a_U$) and assigning a $J_2$ moment to its potential, of magnitude \citep{Burns1976}
\begin{align}
\label{J2}
J_2 =  \frac{1}{2} \sum_{i=1}^{4} \frac{m_i a_i^2}{M \mathcal{R}^2}.
\end{align}
In doing so, we successfully capture the secular perihelion advance generated by the giant planets, without contaminating the results with the effects of close encounters. Any orbit with a perihelion distance smaller than Uranus's semi-major axis was removed from the simulation. Similarly, any particle that came within one Hill radius of the perturber was also withdrawn. The integration time spanned 4 Gyr for each calculation.

As in the case of the analytical model, we constructed six test particle phase-space portraits\footnote{We note that in order to draw a formal parallel between numerically computed phase space portraits and their analytic counterparts, a numerical averaging process must be appropriately carried out over the rapidly varying angles \citep{Morbidelli1993}. Here, we have not performed any such averaging and instead opted to simply project the orbital evolution in the $e-\Delta\varpi$ plane.} in the semi-major axis range $a\in(50,550),$ for each combination of perturber parameters. Each portrait was composed of 40 test particle trajectories whose initial conditions spanned $e\in(0,0.95)$ in increments of $\Delta e = 0.05$ and $\Delta\varpi = 0,180\,\deg$. Mean anomalies of the particles and the perturber were set to 0 and $180\,\deg$ respectively, and mutual inclinations were assumed to be null.

Unlike analytic calculations, here we did not require the perturber's semi-major axis to exceed that of the test particles. Accordingly, we sampled a grid of semi-major axes $a'\in(200,2000)\,$AU and eccentricities $e'\in(0.1,0.9)$ in increments of $\Delta a' = 100\,$AU and $\Delta e'$ = 0.1 respectively. Given the qualitatively favorable match to the data that a $m' = 10 \,m_{\oplus}$ companion provided in the preceding discussion, we opted to retain this estimate for our initial suite of simulations. 

Computed portraits employing the same perturber parameters as before, are shown in Figure (\ref{Fig4}). Drawing a parallel with Figure (\ref{Fig3}), it is clear that trajectories whose secular evolution drives persistent apsidal alignment with the perturber (depicted with orange lines) are indeed reproduced within the framework of direct N-body simulations. However, such orbital states possess minimal perihelion distances that are substantially larger than those ever observed in the real scattered Kuiper belt. Indeed,, apsidally aligned particles that are initialized onto orbits with eccentricities and perihelia comparable to those of the distant Kuiper belt, are typically not long-term stable. Consequently, the process described in the previous section appears unlikely to provide a suitable explanation for the physical clustering of orbits in the distant scattered disk.

While the secular confinement mechanism is disfavored by the simulations, Figure (\ref{Fig4}) reveals that important new features, possessing the same qualitative properties, materialize on the phase-space portrait when the interactions between the test particle and the perturber are modeled self-consistently. Specifically, there exist highly eccentric, low perihelion apsidally \textit{anti-aligned} orbits that are dynamically long-lived (shown with blue dots). Such orbits were not captured by the analytical model presented in the previous section, yet they have orbital parameters similar to those of the observed clustered KBOs.

\begin{figure*}[t]
\includegraphics[width=\textwidth]{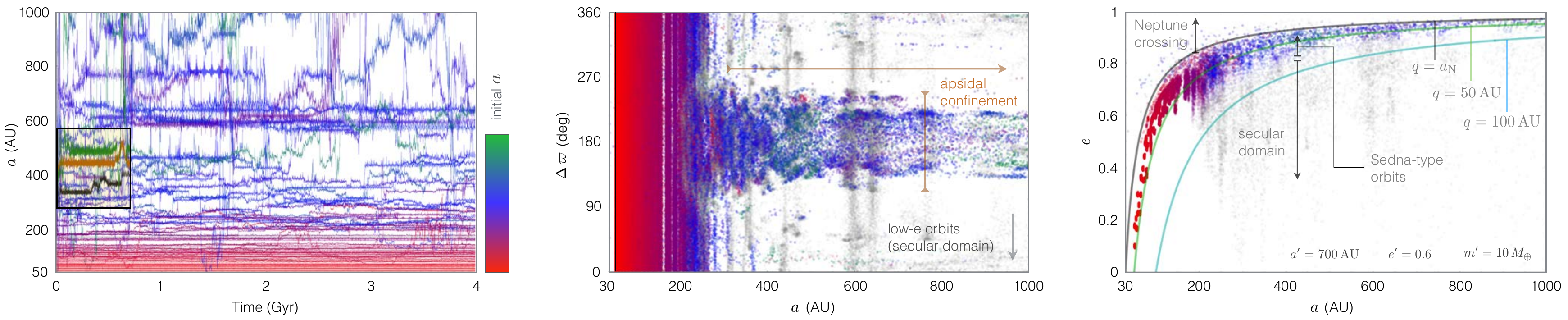}
\caption{A synthetic scattered disk generated with direct N-body simulations. The left panel shows semi-major axis evolution of the dynamically long-lived particles, where color serves as proxy for initial semi-major axis (as shown to the right of the graph). Beyond $a\gtrsim250\,$AU, dynamical evolution is unsteady as the semi-major axes explore a complex web of mean-motion resonances while maintaining stability. The evolution of low-order resonant angles associated with the three boxed, highlighted orbits are shown in Figure (\ref{Fig6}). The middle panel presents the $\Delta\varpi$ footprint generated by the particles as a function of $a$. Point transparency is taken as a proxy observability. Unobservable particles that fall below the observability threshold are shown with gray points. The right panel shows the corresponding $e$ footprint produced by the particles. Note that a subset of KBOs in the simulations are scattered into the secular high-$q$ domain of the phase-space portrait, and constitute a unique prediction of the envisioned perturbation mechanism.}
\label{Fig5}
\end{figure*}

The longevity of highly eccentric apsidally anti-aligned trajectories depicted in Figure (\ref{Fig4}) is surprising, given that they traverse the orbit of the perturber. What is the physical mechanism responsible for their confinement and long-term stability? It is trivial to show that under the assumption of purely Keplerian motion, particles on crossing orbits would experience recurrent close encounters with the perturber over the multi-Gyr integration period. In Figure (\ref{Fig4}), this is made evident by the fact that the narrow region of orbital confinement is surrounded in phase-space by unstable trajectories (shown with gray points). At the same time, it is well known that destabilizing conjunctions can be circumvented via the so-called \textit{phase protection mechanism}, inherent to mean motion commensurabilities \citep{MorbidelliMoons1993,Michtchenko2008}. Accordingly, we hypothesize that the new features observed within the numerically computed phase-space portraits arise due to resonant coupling with the perturber, and the narrowness of the stable region is indicative of the resonance width.

Within the framework of the restricted circular three-body problem (where the perturber's orbit is assumed to be fixed and circular), resonant widths initially grow with increasing particle eccentricity, but begin to subside once the eccentricity exceeds the orbit-crossing threshold \citep{NesvornyRoig2001,RobutelLaskar2001}. Perhaps the situation is markedly different within the context of the highly elliptic restricted three-body problem (as considered here). Specifically, it is possible that even at very high eccentricities, the individual widths associated with the various resonant multiplets remain sufficiently large for a randomly placed orbit to have a non-negligible chance of ending up in resonance with the perturber. We note that associated with each individual resonance is a specific angle (a so-called \textit{critical argument}) that exhibits bounded oscillations. Explicit identification of such angles will be undertaken in the next section.

In an effort to explore the dependence of our results on mass, we constructed two additional suites of phase-space portraits spanning the same semi-major and eccentricity range, with $m' = 1 \,m_{\oplus}$ and $m' = 0.1 \,m_{\oplus}$. Generally, our results disfavor these lower masses. In the instance of an $m' = 0.1 \,m_{\oplus}$ perturber, dynamical evolution proceeds at an exceptionally slow rate, and the lifetime of the solar system is likely insufficient for the required orbital sculpting to transpire. The case of a $m' = 1 \,m_{\oplus}$ perturber is somewhat more promising in a sense that long-lived apdially anti-aligned orbits are indeed evident on the phase-space portraits. However, removal of unstable orbits (i.e. those that reside between the low-perihelion apdally anti-aligned and high-perihelion apsidally aligned stable regions) occurs on a much longer timescale compared with the case of our nominal perturber mass, yielding phase-space portraits that are markedly more contaminated with metastable trajectories, in comparison to those shown in Figure (\ref{Fig4}). Such phase-space portraits likely imply an orbital structure of the Kuiper belt that shows preference for a particular apsidal direction but does not exhibit true confinement, like the data. Accordingly, for the remainder of the paper, we shall adopt $m' = 10 \,m_{\oplus}$ as a representative quantity, keeping in mind the order-of-magnitude nature of this estimate.

\section{Synthetic Scattered Disk} \label{sect5}

Having demonstrated that a massive, distant, eccentric planet can sustain a population of low-perihelion apsidally anti-aligned small bodies against differential precession, we now turn our attention to the question of how the observed population of distant Kuiper belt objects can be produced from a unmethodical starting configuration. To address this inquiry, we have performed an array of numerical experiments aimed at generating synthetic scattered disks. 

\subsection{A Planar Perturber}

Incorporating a series of planar perturber orbits that demonstrated promising phase space portraits, we explored the long-term behavior of a scattered disk population comprised of test-particle orbits whose perihelion orientations were initially randomized. Unlike the previous section, where the presence of the four giant planets was modeled with an enhanced stellar quadrupole field, here all planetary perturbations were accounted for in a direct, N-body manner. The surface density profile of the disk was (arbitrarily) chosen such that each increment of semi-major axes contained the same number of objects on average. Suitably, disk semi-major axes spanned $a\in(50,550)\,$AU as before, with a total particle count of 400. 

The initial perihelion distance was drawn from a flat distribution extending from $q=30\,$AU to $50\,$AU. Additionally, test particle inclinations were set to zero at the start of the simulations, although they were allowed to develop as a consequence of interactions with the giant planets, which possessed their current spatial orbits. As in previous calculations, the system was evolved forward in time for 4 Gyr. 

Remarkably, we found that capture of KBO orbits into long-lived apsidally anti-aligned configurations occurs (albeit with variable success) across a significant range of companion parameters (i.e. $a' \sim 400-1500\,$AU, $e'\sim0.5-0.8$). The characteristic orbital evolution of an evolved synthetic scattered disk corresponding to previously employed perturber parameters is depicted in Figure (\ref{Fig5}). Specifically, the left panel shows the evolution of the semi-major axes of the test particles. Only objects that have remained stable throughout the integration are plotted, and the color scheme is taken to represent the initial semi-major axis. 

Clearly, orbital evolution correspondent to semi-major axes beyond $a\gtrsim 250\,$AU is vastly different from that of the closer-in orbits. While semi-major axes of the inner scattered disk remain approximately constant in time\footnote{We note that this behavior is largely an artifact of the relatively small number of particles initially present in the simulations. Because recurrent close encounters with Neptune dynamically remove KBOs from the scattered disk, the bodies that remain stable for 4 Gyr tend to reside in the so-called extended scattered disk, on orbits that are insulated from short-periodic interactions with Neptune. As a result, such orbits are over-represented in the $a\in(50,250)\,$AU region of Figure (\ref{Fig5}), compared to the real Kuiper belt.} (in line with the secular approximation employed in section \ref{sect2}), semi-major axes of distant Kuiper belt objects skip around over an extensive range, temporarily settling onto distinct values. Such objects experience modulation in both eccentricity and inclination, with a subset even achieving retrograde orbits (we come back to the question of inclinations below). Importantly, this behavior is characteristic of the so-called \textit{Lagrange instability} wherein marginally overlapped mean-motion resonances allow the orbits to diffuse through phase-space, but nevertheless protect them from the onset of large-scale scattering (i.e. a violation of Hill stability - see \citealt{Deck2013} for an in-depth discussion). 

\begin{figure}[t]
\centering
\includegraphics[width=0.9\columnwidth]{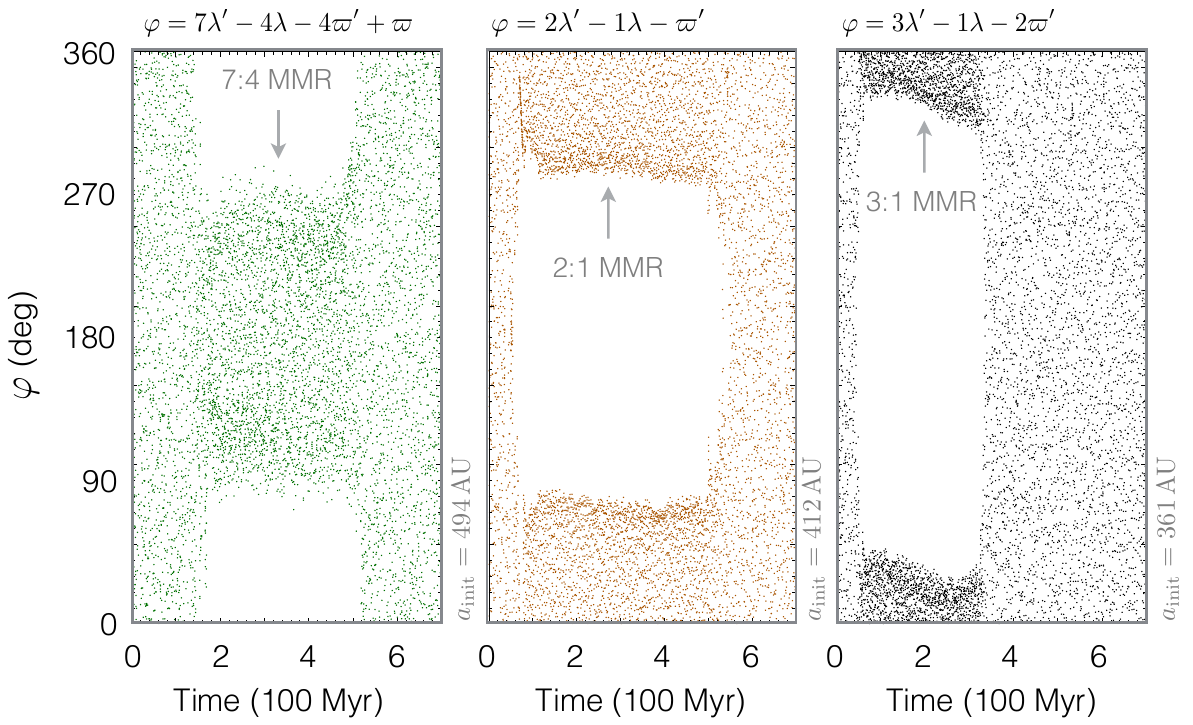}
\caption{Evolution of (comparatively) low-order resonant angles corresponding to the three highlighted semi-major axis time series shown in the left panel of Figure (\ref{Fig5}). The color scheme of the points is identical to that employed within the highlighted box in Figure (\ref{Fig5}). The specific critical argument that enters a temporary mode of libration during the shown timespan is quoted on top of each panel, and the starting semi-major axes are shown on the right of each plot. }
\label{Fig6}
\end{figure}

The observed behavior of the semi-major axes is indicative of the notion that resonant perturbations are responsible for orbital clustering, as discussed above. In an effort to further corroborate the suspicion that resonant perturbations are relevant to the observed high-eccentricity orbits, we have searched for libration of various critical arguments of the form $\varphi = j_1 \lambda + j_2 \lambda' + j_3 \varpi + j_4 \varpi'$ where $\Sigma \, j_i = 0$, within the first Gyr of evolution of a subset of long-term stable orbits with initial semi-major axes beyond $a>250\,$AU. Remarkably, we were able to identify temporary libration of such angles associated with 2:1, 3:1, 5:3, 7:4, 9:4, 11:4, 13:4, 23:6, 27:17,  29:17, and 33:19 mean-motion commensurabilities. Three low-order examples of persistent libration are shown in Figure (\ref{Fig6}), where the color of the simulation data corresponds to that of the boxed, emphasized orbits in the left panel of Figure (\ref{Fig5}). 

While this characterization is emblematic of resonant interactions as a driver for apsidally anti-aligned confinement and enduring stability of test particle orbits, extension of analytical theory into the realm of the unaveraged, highly elliptical three-body problem is coveted for a complete assessment of the dynamical phenomenon at hand. Moreover, we note that while we have exclusively searched for critical angles associated with two-body commensurabilities, it is likely that three-body resonances that additionally include contributions from Neptune's orbital motion, play a significant role in establishing a resonant web along which KBO orbits diffuse \citep{NesvornyRoig2001}. 

Because the semi-major axes of the test particles do not remain fixed in the simulations, their role is best interpreted as that of \textit{tracers} that rapidly explore all parameter space (while remaining on the same resonant web), available within the given dynamical regime. Accordingly, when examining the evolution of the apsidal angle of the KBOs with respect to that of the perturber, it is sensible to plot the entire integration span of the stable subset of orbits. The corresponding footprint of $\Delta\,\varpi$ is shown as a function of $a$ in the middle panel of Figure (\ref{Fig5}). 

The points depicted in this panel vary both in color and transparency. We have used transparency as an approximate proxy for observability: points are rendered progressively more transparent\footnote{Practically, we have chosen to use the Gaussian error function to smoothly connect maximal and minimal transparencies. Obviously, this means of modeling observability is only envisaged as a rough approximation, and a more sophisticated filtering approach based on real survey data can in principle be undertaken.} as perihelion distance increases above $q>30\,$AU and orbital inclination grown closer to $i>40\,\deg$, where an object would be less likely to be detected in a typical ecliptic survey. As before, the color of the points is taken to represent starting semi-major axes, except in the case where transparency is maximized due to the perihelion distance increasing beyond $q>100\,$AU or inclination rising above $i>40\,\deg$. Evolution of objects beyond this observability threshold is shown with nearly transparent gray points. As can be clearly discerned, orbital evolution of stable KBOs with perihelion distances in the observable range are preferentially concentrated in apsidally anti-aligned states. 

It is important to note that not all stable objects within the simulations occupy the (likely) resonant high-eccentricity configurations. That is, there exists an additional population of lower-eccentricity orbits that inhabit the secular domain of the phase-space portrait, and glean long-term stability through apsidal alignment. These objects are primarily represented as gray points in the right panel of Figure (\ref{Fig5}), and constitute a unique, testable consequence of the dynamical mechanism described herein. Specifically, \textit{if a distant, eccentric perturber is responsible for the observed orbital clustering in the distant Kuiper belt, then observational probing of high-perihelion scattered disk with $a>250\,$AU should reveal a collection of objects, whose longitudes of perihelia are on average, $180 \deg$ away from the known objects.}

With an eye towards placing better constraints on $m'$, we have carried out an additional suite of simulations, which confirm that a perturber with mass substantially below our nominal estimate (e.g. $m'=1\,m_{\oplus}$) is unable to generate the degree of orbital clustering seen in the data. Nonetheless, we reiterate that the perturber's elements quoted in Figure (\ref{Fig5}) are not the only combination of parameters that can yield orbital confinement in the distant Kuiper belt. Particularly, even for a fixed value of $m'$, the critical semi-major axis that corresponds to the onset of apsidal clustering depends on $e'$ and $a'$ in a degenerate manner.

A unifying feature of successful simulations that place the transitionary semi-major axis at $a_{\rm{crit}}\sim 250\,$AU, is that the perturber's orbit has a perihelion distance of $q'\sim200-300\,$AU. Approximately mapping $a_{\rm{crit}}$ within our suite of numerical simulations, we empirically find that it roughly follows the relationship
\begin{align}
a_{\rm{crit}} \appropto e'\big(1-\left(e' \right)^2 \big)^{-1} \big(a' \big)^{-2}.
\end{align}
We note however, that the above scaling has limitations: resonant trajectories of the kind shown (with blue dots) in Figure (\ref{Fig4}) only arise in the correct regime at high perturber eccentricities (i.e. $e'\gtrsim0.4-0.5$), and are only stable below $e'\lesssim0.8-0.9$. Analogously, perturber orbits outside of the semi-major axis range $a' = 400 - 1500\,$AU are disfavored by our simulations because parameters required for the onset of orbital clustering at $a\gtrsim250\,$AU lead to severe depletion of the particle population.

\subsection{An Inclined Perturber}

As already discussed in section (\ref{sect2}), an adequate account for the data requires the reproduction of grouping in not only the degree of freedom related to the eccentricity and the longitude of perihelion, but also that related to the inclination and the longitude of ascending node. Ultimately, in order to determine if such a confinement is achievable within the framework of the proposed perturbation model, numerical simulations akin to those reported above must be carried out, abandoning the assumption of coplanarity. Before proceeding however, it is first useful to examine how dynamical locking of the ascending node may come about from purely analytical grounds. 

We have already witnessed in section (\ref{sect3}) that while secular perturbation theory does not adequately capture the full dynamical picture, it provides a useful starting point to guide subsequent development. Correspondingly, let us analyze the dynamical evolution of the $i-\Delta\Omega$ degree of freedom under the assumptions that the relevant equations of motion can be solved in a quasi-isolated fashion\footnote{In adiabatic systems with two degrees of freedom, dynamical evolution of the individual degrees of freedom can proceed in a quasi-decoupled manner (as long as homoclinic curves are not encountered) as a consequence of separation of timescales \citep{Wisdom1983,HenrardCaranicolas1990,BatyginMorbidelli2013}. However, because the system at hand falls outside of the realms of conventional perturbation theory, it is difficult to assert in a a-priori manner if the assumption of decoupled evolution is well justified.} and that unlike the case of $e-\varpi$ dynamics, secular terms dominate the governing Hamiltonian\footnote{Given the tremendous difference in the degrees of excitation of eccentricities and inclinations in the distant scattered belt, it may be plausible to assume that the multiplets of mean motion commensurabilities primarily responsible for maintaining apsidal anti-alignment correspond to \textit{eccentricity} resonances, and only affect orbital inclinations through high-order terms, leaving secular effects to dominate the evolution \citep{EllisMurray2000}.}. Our aim is thus to construct an approximate, but integrable secular normal form that will hopefully capture the dominant mode of spatial angular momentum exchange.

\begin{figure}[t]
\centering
\includegraphics[width=0.85\columnwidth]{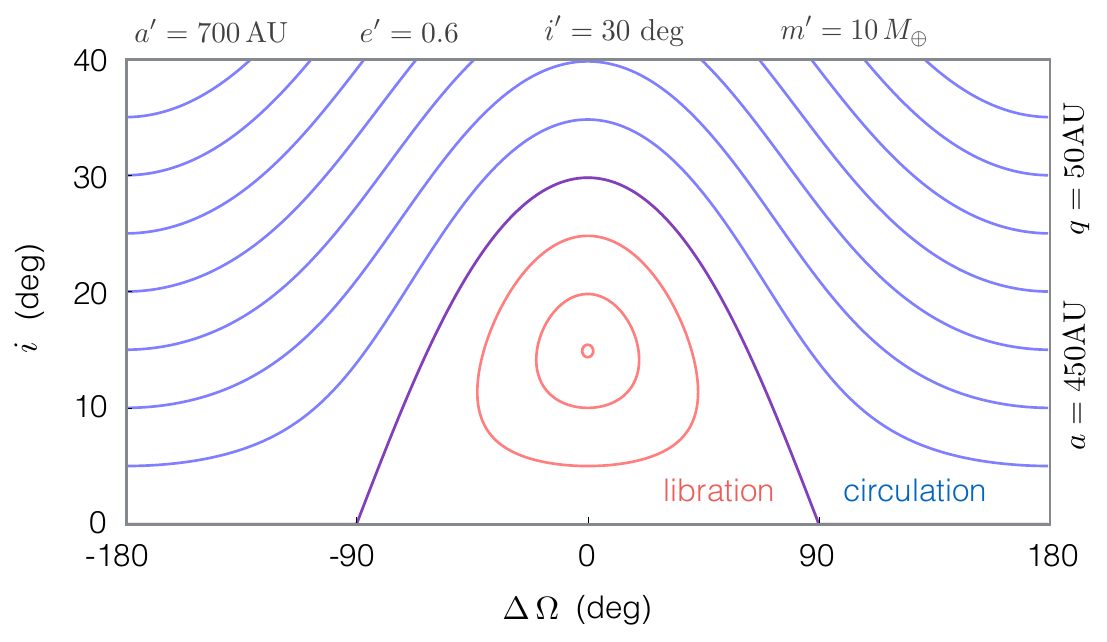}
\caption{Forced secular equilibrium associated with the inclination degree of freedom of the particles. The figure depicts level curves of the Hamiltonian (\ref{Hinc}) for an object with $a=450\,$AU and $q=50\,$AU. As in Figure (\ref{Fig3}), red and blue curves are used to denote librating and circulating trajectories respectively.}
\label{Fig7}
\end{figure}

Following the argument presented in section (\ref{sect3}), the dynamical evolution of the perturber can be delimited to steady regression of the node at constant inclination: $\Omega'=-\mu\,t$. The rate of recession, $\mu$, is equal to the value obtained from equation (\ref{Precession}), diminished by a factor of $\cos(i')$ (e.g. \citealt{SpaldingBatygin2014,Li2014}). Generally speaking, this simplification is not enough to render the secular Hamiltonian integrable, since even at the quadrupole level of approximation, the number of harmonic terms is too great for a successful reduction to a single degree of freedom \citep{Kaula1964,Mardling2010}. Fortunately, however, all objects in the distant Kuiper belt have relatively low orbital inclinations, allowing us to discard harmonics that exhibit dependence on $\sin^{2}(i)$ in favor of the lower-order terms. This refinement yields a quadrupole-level non-autonomous Hamiltonian that contains a single critical argument of the form $(\Omega' - \Omega) = -(\Omega + \mu\,t) = \Delta \Omega$.

Employing a canonical contact transformation generated by the type-2 function $\mathcal{F}_2 = \Psi (-\Omega - \mu\,t)$ (where $\Psi= \sqrt{\G M a} \sqrt{1-e^2}(1-\cos(i))$ is the new action conjugate to $\Delta \Omega$), we obtain an autonomous Hamiltonian whose functional form is reminiscent of equation (\ref{H3}):
\begin{align}
\label{Hinc}
\mathcal{H} &= -\frac{3}{8}\frac{\mathcal{G} M}{a} \frac{\cos(i)}{\left( 1-e^2 \right)^{3/2}} \sum_{i=1}^{4} \frac{m_i a_i^2}{M a^2} \nonumber \\
&-\mu\, \sqrt{\mathcal{G} \, M \, a} \sqrt{1-e^2}  \left(1- \cos(i) \right) \nonumber \\
&- \frac{1}{4} \frac{\mathcal{G}\, m'}{a'} \left( \frac{a}{a'} \right)^2 \big(1-(e')^2\big)^{-3/2} \nonumber \\
&\times \bigg[ \bigg( \frac{1}{4} + \frac{3}{8} e^2\bigg) \left(3 \cos^2(i')-1\right)\left(3 \cos^2(i)-1\right) \nonumber \\
&+ \frac{3}{4}\, \sin(2i') \, \sin(2i) \, \cos(\Delta \Omega) \bigg].
\end{align}
Note that in this Hamiltonian, $e$ takes the form of a parameter rather than a dynamic variable. 

\begin{figure*}[t]
\includegraphics[width=\textwidth]{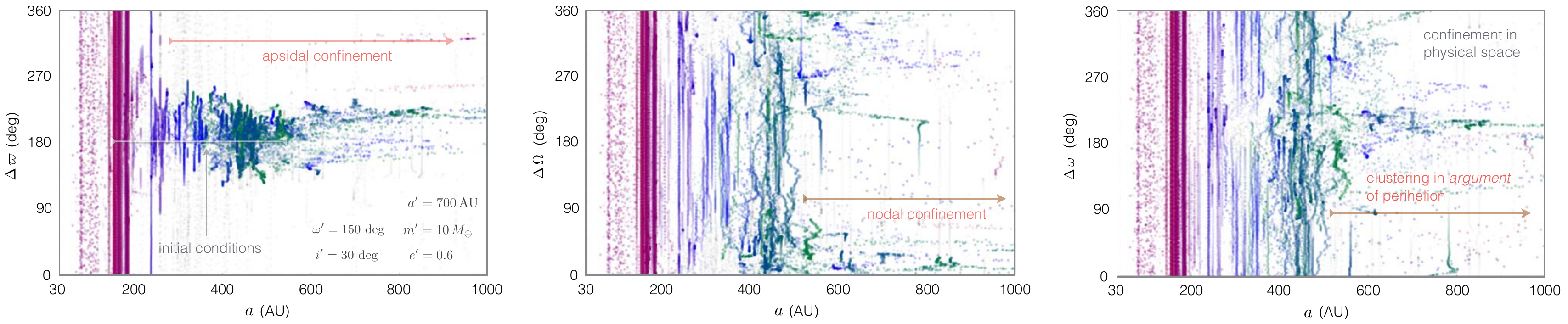}
\caption{The dynamical footprint of a synthetic scattered disk generated within the framework of a simulation where the perturber with the nominal parameters (identical to those employed in Figures \ref{Fig3}-\ref{Fig5}) is taken to reside on an orbit with $i'=30\,\deg$, and initial $\omega' = 150\,\deg$. The left, middle, and right panels depict the longitude of perihelion, longitude of ascending node, and argument of perihelion respectively, as in Figure (\ref{Fig1}). Clearly, alignment of particle orbits in physical space is well reproduced beyond $a>500\,$AU. Note however that unlike Figure (\ref{Fig1}) which simply shows the available data, in this Figure the apsidal and nodal lines are measured with respect to those of the perturber. As in Figure (\ref{Fig5}), color and transparency are used as proxies for starting semi-major axis and observability.}
\label{Fig8}
\end{figure*}

As before, contours of constant $\mathcal{H}$ can be examined as a means of delineating the dynamical flow. Correspondingly, a phase-space portrait with $a = 450\,$AU, $q=50\,$AU, $i'=30\deg$, and other parameters adopted from preceding discussion, is shown in Figure (\ref{Fig7}). Importantly, this analysis demonstrates that a forced equilibrium that facilitates libration of $\Delta \Omega$, exists at $\Delta\Omega=0$ and low particle inclinations. Therefore, to the extent that Hamiltonian (\ref{Hinc}) provides a good approximation to real N-body dynamics, we can expect that a distant perturber can maintain orbital confinement of KBOs, characterized by $\Delta \varpi = 180\deg$ and $\Delta \Omega = 0 \deg$. 

Paired with the observational data (Figure \ref{Fig1}), the theoretical locations of the libration centers provide important clues towards the actual orbit of the perturber. Specifically, if we adopt the statistics inherent to the dynamically stable subset of the clustered population at face value, the simultaneous apsidal anti-alignment and nodal alignment of the perturber with the KBO population implies that $\omega'=138\pm21\deg$. We are now in a position to examine if a scattered KBO population characterized by orbital grouping can be sculpted by the envisioned perturber, with the use of direct N-body simulations. In particular, the following experiments were preformed. 

Similarly to the results shown in Figure (\ref{Fig5}), we constructed a series of synthetic scattered disks. Because we have already demonstrated that (at the relevant eccentricities) the desired apsidal clustering and enduring stability is achieved exclusively for orbits with $\Delta\,\varpi\simeq180\deg$, in this set of simulations the entire scattered disk was initialized with longitudes of perihelion that were anti-aligned with respect to that of the perturber. On the other hand, longitudes of ascending node of the particle orbits were uniformly distributed between $\Omega\in(0,360)\deg$. Inclinations were drawn from a half-normal distribution with a standard deviation of $\sigma_i = 15\deg$, while perihelion distances spanned $q\in(30,50)\,$AU as before. Only objects with initial semi-major axes in the $a\in(150,550)\,$AU range were considered, as previous simulations had shown that dynamical evolution of objects with $a\in(50,150)\,$AU is largely unaffected by the presence of the perturber and is essentially trivial. Correspondingly, each model disk was uniformly populated with $320$ particles. 

In an effort to account for the effects of the giant planets, we adopted a hybrid approach between the averaged and direct methods employed above. Specifically, we mimicked the quadrupolar fields of Jupiter, Saturn, and Uranus by endowing the Sun with a strong $J_2$ moment (given by equation \ref{J2}). Simultaneously, Neptune was modeled in a conventional N-body fashion. This setup allowed for a substantial reduction in computational costs, while correctly representing short-periodic and resonant effects associated with Neptune's Keplerian motion\footnote{In order to ensure that this simplified model captures the necessary level of detail, we have reconstructed the phase-space portraits shown in Figure (\ref{Fig4}) using this setup, and confirmed that the relevant dynamical features are correctly represented.}. As before, each synthetic disk was evolved for $4\,$Gyr. 

For our nominal simulation, we adopted $a'=700\,$AU, $e'=0.6$, and $m'=10\,m_{\oplus}$ as before, and set the inclination and initial argument of perihelion of the perturber to $i'=30\deg$ and $\omega'=150\deg$ respectively. Figure (\ref{Fig8}) shows the simulated confinement of the orbital angles attained in this calculation. Clearly, the results suggest that the clustering seen in the observational data can be reproduced (at least on a qualitative level) by a mildly inclined, highly eccentric distant perturber. Moreover, the libration center of $\Delta \Omega$ indeed coincides with the aforementioned theoretical expectation, suggesting that the dynamical origins of nodal grouping are in fact, secular in nature. Evidently, orbits that experience modulation of orbital inclination due to the circulation of $\Delta\,\Omega$ are preferentially rendered unstable and removed from the system. Drawing a parallel with Figure (\ref{Fig1}), the longitude of perihelion, the longitude of ascending node, and the argument of perihelion are shown in the same order. We note however, that in Figure (\ref{Fig8}) these quantities are measured with respect to the perturber's orbit (without doing so, orbital confinement becomes washed out due to the precession of the perturber's orbit itself). In this sense, the actual values of $\varpi$ and $\Omega$ observed in the data hold no physical meaning, and are merely indicative of the orientation of the perturber's orbit.

\begin{figure*}[t]
\includegraphics[width=\textwidth]{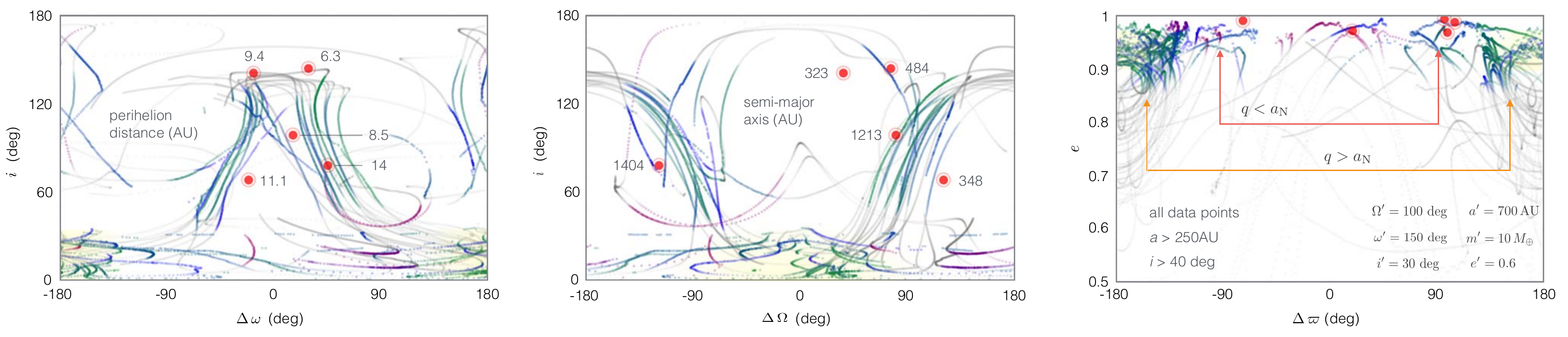}
\caption{High inclination particle dynamics within the synthetic scattered disk. Only trajectories with $a>500\,$AU, corresponding to the physically aligned region of the synthetic disk, are plotted. The left and middle panels show orbital inclination as a function of relative argument of perihelion and relative longitude of ascending node respectively. The clustered low-inclination populations are highlighted with a local yellow background. Although the high-inclination component of the dynamical footprint is not shown in Figure (\ref{Fig3}), here it is clear that it exhibits coherent structure and that initially low-inclination objects can acquire extreme inclinations as a result of interactions with the perturber. Real Kuiper belt objects with $a>250\,$AU and $i>40\,\deg$ are shown as red points, where the data has been appropriately translated assuming $\omega'=140\deg$ and $\Omega'=100\,\deg$, as inferred from Figure (\ref{Fig1}). Numbers quoted next to the points denote perihelion distance and semi-major axis of the data on the left and middle panels respectively. The right panel shows the eccentricity as a function of the relative longitude of perihelion. Evidently, maximal eccentricity is attained away from exact apsidal anti-alignment, consistent with the exceptionally low perihelion distances associated with the existing dataset.}
\label{Fig9}
\end{figure*}

Although the clustering of the simulation data in Figure (\ref{Fig8}) is clearly discernible, an important difference comes into view, when compared with the planar simulation portrayed in Figure (\ref{Fig5}). In the case of an inclined perturber, clustering in $\Omega$ is evident only beyond $a\gtrsim500\,$AU, while the same clustering in $\varpi$ appears for $a\gtrsim250\,$AU, as in Figure (\ref{Fig5}). This discrepancy may be indicative of the fact that an object somewhat more massive than $m'=10\,m_{\oplus}$ is required to shift the dividing line between randomized and grouped orbits to smaller particle semi-major axes. Additionally, apsidal confinement appears substantially tighter in Figure (\ref{Fig8}), than its nodal counterpart. To this end, however, we note that the initial values of the particles' longitudes of perihelia were chosen systematically, while the observed nodal clustering has been dynamically sculpted from an initially uniform distribution. This difference may therefore be an artifact of the employed initial conditions. 

For completeness, we performed an additional suite of numerical integrations, varying the inclination of the perturber within the $i'\in(60,180)\deg$ range, in increments of $\Delta\,i'=30\deg$. For each choice of inclination, we further iterated over the perturber's argument of perihelion $\omega'\in(0,360)\deg$ with $\Delta\,\omega'=30\deg$, retaining the initial longitude of ascending node at the same (arbitrarily chosen) value and adjusting the initial value of $\varpi$ of the scattered disk objects accordingly. This set of calculations generally produced synthetic scattered disks that were less reminiscent of the observational data than our nominal calculation, further suggesting that the distant perturber likely resides on a low-inclination orbit, with an argument of perihelion a few tens of degrees below $180\,\deg$.

Although Figure (\ref{Fig8}) only emphasizes objects with inclinations below $i\leqslant40\deg$, particles within simulations that feature an inclined perturber generally explore highly oblique orbits as well. The evolutionary tracks of such objects, projected onto a $i-\Delta\,\omega$, $i-\Delta\,\Omega$, and $e-\Delta\varpi$ planes are shown in Figure (\ref{Fig9}), where point transparency is taken to only indicate the perihelion distance. From these illustrations, it is evident that conventional members of the distant scattered disk population may disappear from view due to an increasing perihelion distance with growing inclination, only to subsequently reappear on misaligned and highly inclined orbits. This form of orbital evolution is likely associated with Kozai dynamics \textit{inside} mean-motion resonances (see Ch. 11 of \citealt{MorbyBook}), and depends weakly on the inclination of the perturber. 

Such results are indeed suggestive, as a small number of highly inclined objects (whose origins remain elusive) does indeed exist within the observational census of the Kuiper belt. Specifically, known KBOs with $a>250\,$AU and $i>40\,\deg$ are overplotted in Figure (\ref{Fig9}) as red dots\footnote{Note that these objects do not appear in Figure (\ref{Fig1}) because they have $q<30\,$AU.}. The agreement between the theoretical calculation and data is more than satisfactory, and is fully consistent with the recent analysis of \citet{Gomes2015}, who also analyzed this population and concluded that it can be best explained by the existence of a distant planet in the extended scattered disk. Astonishingly, our proposed explanation for orbital clustering signals unexpected consistency with a superficially distinct inferred population of objects that occupy grossly misaligned orbits. Therefore, if a distant perturber of the kind considered in this work is truly responsible for the observed structure of the Kuiper belt, continued characterization of the high-inclination component of the scattered disk may provide an indirect observational handle on the orbital parameters of the perturbing body.  

We end this section by drawing attention to the fact that while numerical construction of synthetic scattered disks presented here has been of great utility, these simulations are not fully realistic. That is, although in this work we have adopted the current giant planet orbits for definitiveness, the actual process of Kuiper belt formation was likely associated with initially eccentric and inclined giant planet orbits, that subsequently regularized due to dynamical friction \citep{Tsiganis2005,Levison2008,Batygin2011,Nesvorny2015}. This means that initial assembly of the clustered population could have been affected by processes that no longer operate in the present solar system. As a consequence, extension of the reported numerical simulations to account for self-consistent formation of the Kuiper belt likely constitutes a fruitful avenue to further characterization of the proposed perturbation model. 

\section{Summary} \label{sect6}

To date, the distinctive orbital alignment observed within the scattered disk population of the Kuiper belt, remains largely unexplained. Accordingly, the primary purpose of this study has been to identify a physical mechanism which can generate, and maintain the peculiar clustering of orbital elements in the remote outskirts of the solar system. Here, we have proposed that the process of \textit{resonant coupling} with a distant, planetary mass companion can explain the available data, and have outlined an observational test that can validate or refute our hypothesis. 

We began our analysis with a reexamination of the available data. To this end, in addition to the previously known grouping of the arguments of perihelia \citep{TrujilloSheppard}, we have identified ancillary clustering in the longitude of the ascending node of distant KBOs, and showed that objects that are not actively scattering off of Neptune exhibit true orbital confinement in inertial space. The aim of subsequent calculations was then, to establish whether gravitational perturbations arising from a yet-unidentified planetary mass body that occupies an extended but nevertheless bound orbit, can adequately explain the observational data.

The likely range of orbital properties of the distant perturber was motivated by analytic considerations, originating within the framework of octupole-order secular theory. By constructing secular phase-space portraits of a strictly planar system, we demonstrated that a highly eccentric, distant perturber can drive significant modulation of particle eccentricities and libration of apsidal lines, such that the perturber's orbit continuously encloses interior KBOs. Intriguingly, numerical reconstruction of the projected phase-space portraits revealed that in addition to secular interactions, resonant coupling may strongly affect the dynamical evolution of KBOs residing within the relevant range of orbital parameters. More specifically, direct N-body calculations have shown that grossly overlapped, apsidally anti-aligned orbits can be maintained at nearly Neptune-crossing eccentricities by a highly elliptical perturber, resulting in persistent near-colinearity of KBO perihelia.

Having identified an illustrative set of orbital properties of the perturber in the planar case, we demonstrated that an inclined object with similar parameters can dynamically carve a population of particles that is confined both apsidally and nodally. Such sculpting leads to a family of orbits that is clustered in physical space, in agreement with the data. Although the model proposed herein is characterized by a multitude of quantities that are inherently degenerate with respect to one another, our calculations suggest that a perturber on a $a'\sim700\,$AU, $e'\sim0.6$ orbit would have to be somewhat more massive (e.g. a factor of a few) than $m'= 10\,m_{\oplus}$ to produce the desired effect. 

A unique prediction that arises within the context of our resonant coupling model is that the perturber allows for the existence of an additional population of high-perihelion KBOs that \textit{do not} exhibit the same type of orbital clustering as the identified objects. Observational efforts aimed at discovering such objects, as well as directly detecting the distant perturber itself constitute the best path towards testing our hypothesis. 

\section{Discussion} \label{sect7}

The resonant perturbation mechanism proposed herein entails a series of unexpected consequences that successfully tie together a number of seemingly unrelated features of the distant Kuiper belt. In particular, the long-term modulation of scattered KBO eccentricities provides a natural explanation for the existence of the so-called distant detached objects such as Sedna and 2012$\,$VP113 \citep{Brown2004,TrujilloSheppard}. Viewed in this context, the origins of such bodies stem directly from the conventional scattered disk, and should on average exhibit the same physical characteristics as other large members of the Kuiper belt. Moreover, we note that because these objects are envisioned to chaotically explore an extensive network of mean motion resonances, their current semi-major axes are unlikely to be indicative of their primordial values.

Another unanticipated result that arises within the context of our narrative is the generation of a highly inclined population of orbits. \citet{Gladman2009} has suggested that the presence of highly inclined KBOs such as Drac point to a more extensive reservoir of such bodies within the Kuiper belt. Not only is our proposed perturbation mechanism consistent with this picture, it further implies that this population is inherently connected to the scattered disk. Accordingly, the dynamical pathway toward high inclinations should become apparent through observational characterization of high-perihelion objects that also exhibit substantial eccentricities. 

As already alluded to above, the precise range of perturber parameters required to satisfactorily reproduce the data is at present difficult to diagnose. Indeed, additional work is required to understand the tradeoffs between the assumed orbital elements and mass, as well as to identify regions of parameter space that are incompatible with the existing data. From an observational point of view (barring the detection of the perturber itself), identification of the critical eccentricity below which the observed orbital grouping subsides may provide important clues towards the dynamical state of the perturber. Simultaneously, characterization of the aforementioned high-inclination population of KBOs may yield meaningful constraints on the perturber's orbital plane. 

Although our model has been successful in generating a distant population of small bodies whose orbits exhibit alignment in physical space, there are observational aspects of the distant Kuiper belt that we have not addressed. Specifically, the apparent clustering of arguments of perihelia near $\omega \sim 0$ in the $a\sim150-250\,$AU region remains somewhat puzzling. Within the framework of the resonant perturber hypothesis, one may speculate that in this region, the long-term angular momentum exchange with the planets plays a sub-dominant, but nevertheless significant role, allowing only critical angles that yield $\varpi-\Omega = \omega \sim 0$ to librate. Additional calculations are required to assess this presumption. 

Another curious feature of the distant scattered disk is the lack of objects with perihelion distance in the range $q=50-70\,$AU. It is yet unclear if this property of the observational sample can be accounted for, by invoking a distant eccentric perturber such as the one discussed herein. Indeed, answering these questions comprises an important avenue towards further characterization of our model.

In this work, we have made no attempt to tie the existence of a distant perturber with any particular formation or dynamical evolution scenario relevant to the outer solar system. Accordingly, in concluding remarks of the paper, we wish to briefly speculate on this subject. With an eye towards producing a planet akin to the one envisioned by \citet{TrujilloSheppard}, \citet{KenyonBromley2015} have argued for the possibility of forming a Super-Earth type planet at $a\sim150-250\,$AU over the course of the solar system's lifetime. While observational inference of extrasolar planetary systems such as HR$\,$8799 \citep{Marois2008} suggests that planets can indeed occupy exceptionally wide orbits, the solar nebula would have had to be exceptionally expansive to be compatible with \textit{in-situ} formation of a planet on such a distant and eccentric orbit, as the one considered here.

Instead of the \textit{in-situ} hypothesis, our proposed perturber may be more reasonably reconciled with a dynamical scattering origin. Specifically, it is possible that our perturber represents a primordial giant planet core that was ejected during the nebular epoch of the solar system's evolution. Recent simulations have demonstrated that such a scenario may in fact be an expected outcome of the early evolution of planetary systems \citep{BromleyKenyon2014}. Moreover, the calculations of \citet{Izidoro2015}, aimed at modeling the formation of Uranus and Neptune through a series of giant impacts (needed to reproduce the planetary obliquities - see e.g. \citealt{Morby2012}), have demonstrated that a system of protoplanetary cores typically generates more than two ice-giant planets. Accordingly, the work of \citet{Izidoro2015} predicts that one or more protoplanetary cores would have been ejected out of the solar system. Within the context of this narrative, interactions with the Sun's birth cluster, and possibly the gaseous component of the nebula, would have facilitated the retention of the scattered planet on a bound orbit.
\\
\\
\textbf{Acknowledgments}  \\ 
We are thankful to Kat Deck, Chris Spalding, Greg Laughlin, Chad Trujillo, and David Nesvorn{\'y} for inspirational conversations, as well as to Alessandro Morbidelli for providing a thorough review of the paper, which led to a substantial improvement of the manuscript.

\end{document}